\newcommand{\beq}[1]{\begin{equation} \label{#1}}
\newcommand{\eeq}{\end{equation}}
\newcommand{\beqa}[1]{\begin{eqnarray} \label{#1}}
\newcommand{\eeqa}{\end{eqnarray}}
\newcommand{\beqn}{\begin{eqnarray*}}
\newcommand{\eeqn}{\end{eqnarray*}}

\newcommand{\dt}[2]{\frac{\partial^{#1}{#2}}{\partial{t}^{#1}}}
\newcommand{\rf}[1]{(\ref{#1})}

\newcommand{\veps}{\varepsilon}
\newcommand{\ga}{\alpha}
\newcommand{\gb}{\beta}

\newcommand{\vphi}{\varphi}

\newcommand{\Dp}[1]{\frac{\partial}{\partial{#1}}}

\documentstyle[12pt,fleqn]{article}

\newcommand{\kap}{\frac{\kappa}{4\pi{e}^2}}
\newcommand{\kpi}{\frac{\kappa}{4\pi}}

\begin{document}
\begin{picture}(174,8)
   \put(31,8){\shortstack[c]
       {RUSSIAN GRAVITATIONAL SOCIETY \\
       ULYANOVSK STATE UNIVERSITY        }       }
\end{picture}
\begin{flushright}
                                         RGS-USU-97/03 \\
                                         gr-qc/9706043    \\
\end{flushright}
\bigskip

\begin{center}
A NEW CLASS OF INHOMOGENEOUS COSMOLOGICAL MODELS WITH YANG-MILLS FIELDS
\end{center}
\begin{center}
V.K.Shchigolev, ~V.M.Zhuravlev,~S.V.Chervon
\end{center}

\begin{center} Ul'yanovsk State University\\
          432700 Ul'yanovsk, Russia \\
          e-mail: shchigol@themp.univ.simbirsk.su\\
          PACS-94: 04.20.Jb \ 04.40.-b
\end{center}
{\small
Exact solutions corresponding to spherically symmetric inhomogeneous
nonstationary Tolman metrics are obtained for the self-consistent
system of Einstein-Yang-Mills equations for the gauge group $SO_3$.
}
\bigskip

The static configurations of the Einstein-Yang-Mills (EYM) system
have been studied in a large number of works (see, for example,
 \cite{C1,C2}). Investigations of the EYM system in application to
cosmological models of the Universe have been much less successful,
and all the basic results in this direction have been obtained for
Friedmann-Robertson -Walker (FRW) spaces. For example, in
\cite{C3,C4} exact solutions of the Yang-Mills (YM) equations were
obtained on a background of homogeneous and isotropic Friedmann metrics.
In \cite{C5}, exact self-consistent solutions of the EYM equations
were found with the aid of a conformally flat representation of a line
element and the conformal invariance of massless YM fields.
However, analysis of these results and additional investigations show
that narrowing the class of the desired spherically symmetric solutions
of the EYM system to homogeneous Friedmann models severely limits the
possibility of understanding the role of YM fields in cosmological processes,
especially during the epoch of the very early Universe \cite{C6}.
It is obvious that any point source of YM fields should destroy the
homogeneity of space. The question of a possible substantial role
of YM fields in cosmological inflation as a result of their strong
nonlinearity was discussed in \cite{C7} from a general standpoint
(irrespective of the class of metrics). Accordingly, to clarify all
basic aspects of the effect of YM fields on the evolution of the Universe,
the admissible forms of the metrics must be extended as much as possible.
On these grounds, in the present paper we propose to abandon the requirement
of homogeneity of space (retaining its isotropy) in order to investigate
how self-gravitating YM fields can influence the regimes of inflation of
the Universe and whether or not in this case the structure of space can
asymptotically in time reach the structure of homogeneous isotropic Friedmann
spaces, as is observed in the present epoch.

We shall study the EYM system on the basis of the $SO_3$ non-Abelian gauge
theory, described by the action.
\beq{SEYM}
    S=-\frac{1}{8\pi}\int\limits_{{\sf\bf M}}\sqrt{-g}d^4x\left\{\frac{{\cal R}}{2\kappa}
    +\frac{1}{2}F^a_{\mu\nu}F_a^{\nu\mu}\right\}
\eeq
Here ${\sf \bf M}$ is the pseudo-Riemannian spase-time with signature
$(+ - - -)$, $g=\det(g_{\ga\gb})$ is the determinant of the metric tensor,
${\cal R}$ is the scalar curvature,
\beq{E2}
F^a_{\mu\nu}=\partial_{\mu}W_\nu^a-\partial_{\nu}W_\mu^a+
ief_{bc}^{a}W^b_{\mu}W^c_{\nu}
\eeq
$W^a_{\mu}$ is the isotriplet of YM fields in the $SO_3$ model,
$e$ is a characteristic constant, $f_{bc}^a$ are the structure constants
of the  $SO_3$ group, and $\kappa=8\pi G$.
The Euler-Lagrange equations for the action  \rf{SEYM} have the form
\beqa{E3}
 &&G^\ga_\gb={\cal R}_\gb^\ga-\frac{1}{2}\delta_\gb^\ga{\cal R}=\kappa T^\ga_\gb,\\
 &&D_\gb(\sqrt{-g}F^{a\ga\gb})=0
\eeqa
where ${\cal R}_\gb^\ga$ is the Ricci tensor, $G_{\ga}^{\gb}$ is the Einstein tensor,
$D_\gb$ is the covariant derivative, and the energy-momentum tensor of the
YM fields equals
\beq{E5}
 T_{\gb}^{\ga}=\frac{1}{4\pi}\left(F^{a\ga\mu}F^{a}_{\gb\mu}
 -\frac{1}{4}\delta_{\gb}^{\ga}F^{a\mu\nu}F^{a}_{\mu\nu}\right)
\eeq

Spherically symmetric cosmological models are described by metrics for which
the spase-time interval has the following general form:
\beq{E6}
 ds^2=dt^2-U(r,t)dr^2-V(r,t)d\Omega^2, \quad d\Omega^2=d\theta^2+\sin^2\theta d\phi^2
\eeq
written in a comoving synchronous coordinate system. For YM fields, the general
spherically symmetric ansatz can be written in the form
\beqa{E7}
&&W_i^a=\veps_{iab}x^b\frac{K(r,t)-1}{er^2}+\delta_{i}^a\frac{S(r,t)}{er}
+x^a{x_i}\frac{T(r,t)}{er},\\ \nonumber
&&W^a_0=x^a\frac{W(r,t)}{er},\quad \Phi^a=x^a\frac{H(r,t)}{er}, \quad T(r,t)=-\frac{S(r,t)}{r^2}
\eeqa
Here and below $\mu,\nu=0,\ldots,3,~i,j=1,2,3,~a,b=1,2,3$; the functions
$K,S,T,W,H$ are unknown.

Introducing the orthonormalized isoframe
\beqa{Rep} \nonumber
&&{\bf n}=(\sin\theta\cos\vphi,\sin\theta\sin\vphi,\cos\theta),\\
&&{\bf l}=(\cos\theta\cos\vphi,\cos\theta\sin\vphi,-\sin\theta),\\
\nonumber
&&{\bf m}=(-\sin\vphi,\cos\vphi,0)
\eeqa
and transforming to spherical coordinates, we find that the
ansatz \rf{E7} becomes
\beqa{E8} \nonumber
&&{\bf W}_1=0\\  \nonumber
&&{\bf W}_2=e^{-1}\left\{(K-1){\bf m}+S{\bf l}\right\},\\
&&{\bf W}_3=e^{-1}\left\{-(K-1){\bf l}+S{\bf m}\right\},\\ \nonumber
&&{\bf W}_0=e^{-1}W{\bf n}.
\eeqa
As a result, we have
\beqa{E9} \nonumber
&& {\bf F}_{01}=-{\bf F}_{10}=-e^{-1}W'{\bf n},\\ \nonumber
&& {\bf F}_{02}=-{\bf F}_{20}=
   e^{-1}\left((\dot{K}+WS){\bf m}+(\dot{S}-WK){\bf l}\right),\\ \nonumber
&& {\bf F}_{03}=-{\bf F}_{30}=
   e^{-1}\left(-(\dot{K}+WS){\bf l}+(\dot{S}-WK){\bf m}\right)\sin\theta\\
&& {\bf F}_{12}=-{\bf F}_{21}=
   e^{-1}\left(K'{\bf m}+S'{\bf l}\right),\\ \nonumber
&& {\bf F}_{23}=-{\bf F}_{32}=
   e^{-1}\sin\theta\left(K^2-1+S^2\right){\bf n},\\ \nonumber
&& {\bf F}_{13}=-{\bf F}_{31}=
   e^{-1}\sin\theta\left(-K'{\bf l}+S'{\bf m}\right)
\eeqa
Here and below an overdot denotes the derivative $\partial/\partial t$ and
the prime denotes the derivative $\partial/\partial r$
To analyze the system of EYM equations, it is also convenient to introduce
the new fields $A(r,t)$, $B(r,t)$ and $\phi(r,t)$ with the aid of the formulas
\beq{E10}
K(r,t)=A(r,t)\sin\phi(r,t), ~S(r,t)=A(r,t)\cos\phi(r,t),
    ~B(r,t)=\dot\phi(r,t)+W(r,t)
\eeq

After the corresponding transformations, the EYM system acquires the following
form
\beqa{E11} \nonumber
&&G_0^0=\kappa T_0^0=
\kap\left[\frac{{W'}^2}{2U}+\frac{\dot{A}^2+A^2B^2 }{V}+
         \frac{{A'}^2+A^2{\phi'}^2}{UV}+\frac{(A^2-1)^2}{2V^2}\right];\\
\nonumber
&&G_1^1=\kappa T_1^1=
\kap\left[\frac{{W'}^2}{2U}-\frac{\dot{A}^2+A^2B^2 }{V}-
         \frac{{A'}^2+A^2{\phi'}^2}{UV}+\frac{(A^2-1)^2}{2V^2}\right];\\
&&G_2^2=G_3^3=\kappa T_2^2=\kappa T_3^3=
\kap\left[-\frac{{W'}^2}{2U}-\frac{(A^2-1)^2}{2V^2}\right];\\ \nonumber
&&G_0^1=\kappa T_0^1=\kap\left[2\frac{A'\dot{A}+A^2B\phi'}{V}\right];\\
\nonumber
&&\Dp{t}\left(\sqrt{U}\dot{A}\right)-\Dp{r}\left(\frac{1}{\sqrt{U}}A'\right)+
\sqrt{U}\left(\frac{(\phi')^2}{U}-B^2+\frac{A^2-1}{V}\right)A=0\\
\label{E12}
&&\Dp{t}\left(\sqrt{U}AB\right)-\Dp{r}\left(\frac{1}{\sqrt{U}}A{\phi}'\right)+
\sqrt{U}\dot{A}B-\frac{1}{\sqrt{U}}A'{\phi}'=0\\  \nonumber
&&\Dp{r}\left(\frac{V}{\sqrt{U}}W'\right)-2\sqrt{U}A^2B=0
\eeqa
where $G_{\gb}^{\ga}$ are known expressions obtained for the metric \rf{E6}.

We now consider a class of exact solutions of the EYM equations that corresponds
to the reduction $A(r,t)=0$, which is equivalent to the condition
$$
    K(r,t)=S(r,t)=0
$$
This requirement does not mean that there is no YM field. As is clear from
\rf{E8}, the nonzero components of the YM fields are as follows:
$$
    {\bf W}_0=e^{-1}W{\bf n},\quad {\bf W}_2=-e^{-1}{\bf m},\quad
    {\bf W}_3=e^{-1}{\bf l}
$$
then the nonzero components of the stress tensor are
\beqa{E14}
&&{\bf F}_{01}=-{\bf F}_{10}=-e^{-1}W'{\bf n},\\ \nonumber
&& {\bf F}_{23}=-{\bf F}_{32}=
   e^{-1}\sin\theta\left(K^2-1+S^2\right){\bf n},
\eeqa

The EYM system  \rf{E11},\rf{E12} with $A\equiv 0$ is highly simplified.
We call attention to the fact that in this case
$$
    G_0^1\equiv \frac{\dot{V'}}{V}-\frac{V'\dot{V}}{2V^2}-\frac{V'\dot{U}}{2UV}=0
$$
This equation can be solved and yields a relation between the functions $U$ É $V$:
$\sqrt{U}=(\sqrt{V})'/f(r)$, where $f(r)$ is an arbitrary functions of $r$.
This class of metrics comprises the well-known Tolman metrics \cite{C11}
with the interval
\beq{E15}
      ds^2=dt^2-\frac{(R')^2}{f^2}dr^2-R^2d\Omega^2,
\eeq
where $R=R(r,t) > 0 $is a function to be determined. For example, for
$R(r,t)=a(t)g(r), f(r)=g'(r), g(r)=\{\sin r, r, {\rm sh}r \}$ the Tolman metric
is identical to the Friedmann metric. In the general case, the Tolman metrics
correspond to inhomogeneous cosmological models.

The following expressions for the Einstein tensor can be derived for the
interval \rf{E15}:
\beq{E16}
 G^0_0= \frac{F'}{2R'R^2},
 ~G^1_1=\frac{\dot{F}}{2\dot{R}R^2},
 ~G^2_2=G^3_3=\frac{1}{4R'R}\left(\frac{\dot{F}}{\dot{R}}\right)'
\eeq
where
\beq{E17}
  F(r,t)=2R\dot{R}^2+2R(1-f^2)
\eeq
The following equations follow from the conservation law ${T^\gb_\ga}_{;\gb}=0$
written in the Tolman metric \rf{E15}:
\beq{E18}
  T^2_2=T^1_1+\frac{R}{2R'}\left(T^1_1\right)',
\eeq
\beq{E19}
     R^2\{ \dot{(T^0_0)}R'-(T^1_1)'\dot{R}\}+(T^0_0-T^1_1)\dt(R^2R')=0
\eeq

With allowance for relations \rf{E15}-\rf{E19}, Eqs. \rf{E12} reduce to one
equation, from which it follows that
\beq{E20}
  W'=eq(t)\frac{R'}{fR^2}
\eeq
where $q(t)$ is an arbitrary function of $t$. We now find from Eqs. \rf{E11}
\beq{E21}
    T_0^0=T_1^1=-T_2^2=-T_3^3=(8\pi)^{-1}Q^2R^{-4},
\eeq
where $Q^2=q^2+g^2$, $g=e^{-1}$. Substituting the components of the
energy-momentum tensor into Eq. \rf{E18} transforms this equation into an
identity for the arbitrary function $R(r,t)$, and Eq. \rf{E19} leads only to
the requirement $Q=const$, which is equivalent to the requirement $q=const$.

After the calculations are performed and Eq.\rf{E16}is taken into account,
the Einstein equations \rf{E11} reduce to three equations:

\beq{E23}
F'=\kpi Q^2 \frac{R'}{R^2}, ~\dot{F}=\kpi Q^2 \frac{\dot{R}}{R^2},
~\left(\frac{\dot{F}}{\dot{R}}\right)'=-2\kpi Q^2 \frac{\dot{R}}{R^3},
\eeq
the last of which follows from the first two. The first two equations, however,
lead to one equation for the function $R(r,t)$:
\beq{E24}
   (R\dot{R})^2=(f^2-1)R^2+\delta R - GQ^2
\eeq
Just as in the standard Tolman model \cite{Tolm}, the solutions can be divided
into three separate classes in accordance with  the conditions:
$f^2 =  1, ~f^2 > 1, ~f^2 <1 $.

1. \underline{Parabolic model $(f^2 = 1)$}. In this case, the solutions  for
$R(r,t)$ can be represented  in the form
\beq{E25}
R(r,t)=R^p(r,t)=\delta^{-1}GQ^2
+\delta^{-1}\left[X^{1/3}_{+}(r,t)+X^{1/3}_{-}(r,t)\right]^2,
\eeq
where $\delta=const>0$
$$
   X_{\pm}(r,t)=Ht-\gb(r)\pm\sqrt{G^3Q^6+(Ht-\gb({r}))^2},
   ~H=\pm3\delta^2/4
$$
Here and below, $\gb(r)$ is an arbitrary differentiable function of $r$.

2. \underline{Hyperbolic model $(f^2 > 1)$}. The general integral of Eq.
\rf{E24} for $R=R^h(r,t)$ can be written in the form
\beqa{E26} \nonumber
&&t-\gb(r)=(f^2-1)^{-1}\left[(f^2-1)R^2+\delta R - G Q^2\right]^2-\\
&&-\frac{\delta}{2}(f^2-1)^{-3/2}{\rm ln}\left\{2(f^2-1)^{1/2}
\left[(f^2-1)R^2+\delta R - G Q^2\right]^{1/2}+2(f^2-1)R+ \delta\right\}
\eeqa
In the particular case $\delta=0$ the solution for $R$ can be found in explicit
form. Specifically,
\beq{E27}
 R(r,t)=R^h_0(r,t)=(f^2-1)^{1/2}\left[(f^2-1)(t-\gb(r))^2+G Q^2\right]^{1/2}.
\eeq

3. \underline{Elliptical model $(f^2 < 1)$}.  A real solution for
$R(r,t)=R^e(r,t)$ exists, if
$$
    \delta >0, ~\delta^2\le 4GQ^2, ~1-\delta^2(4GQ^2)^{-1}< f^2 < 1.
$$
In this case the general integral can be written in the form
\beqa{E28} \nonumber
&&t-\gb(r)=(1-f^2)^{-1}\left[-(1-f^2)R^2+\delta R - G Q^2\right]^2+\\
&&+\frac{\delta}{2}(1-f^2)^{-3/2} {\rm arcsin}\left\{
\frac{\delta-2(1-f^2)R}{\left[(1-f^2)R^2+\delta R - G Q^2\right]^{1/2}}\right\}
\eeqa

The solutions obtained correspond to a space-time filled with a YM field
possessing only a radial electric component  ${\bf\cal E}_r$ and only a radial
magnetic component ${\bf\cal B}_r$ which have the form
\beqa{E}  \nonumber
 {\cal E}_r={\bf F_{r0}}=q\frac{R'}{fR^2}{\bf n},
 ~{\cal B}_r=-\frac{1}{2}\sqrt{q^{*}}\veps_{rjk}{\bf F^{jk}}=g\frac{R'}{fR^2}{\bf n},
\eeqa
where $g^*={\rm det}(g_{ij})=R^4R'^2\sin^2\theta/f^2(r)$. Hence one can see
that the constants $q$ and $g$ have the meaning of electric and magnetic
charges, respectively.

 This work was performed under the partial financial support of the
Scientific Center "Kosmion" as part of the project "Kosmomikrofizika".

\end{document}